\documentclass[aps,pra,superscriptaddress,amsmath,amssymb,showpacs,twocolumn,notitlepage]{revtex4}
\usepackage{amsmath,graphicx,amssymb,epsfig,dsfont,color,subfigure}

\renewcommand{\Im}{{\rm Im}}

\newcommand{\rd}{{\rm d}}

\newcommand{\kb}{k_{\rm B}}
\newcommand{\rs}{{\rm s}}
\newcommand{\rp}{{\rm p}}
\newcommand{\re}{{\rm e}}

\graphicspath{{./Figures/}}

\begin{document}

\title{Hyperbolic waveguide for long-distance transport of near-field heat flux}

\author{Riccardo Messina}
\affiliation{Laboratoire Charles Coulomb (L2C), UMR 5221 CNRS-Universit\'{e} de Montpellier, F- 34095 Montpellier, France}

\author{Philppe Ben-Abdallah}
\affiliation{Laboratoire Charles Fabry,UMR 8501, Institut d'Optique, CNRS, Universit\'{e} Paris-Sud 11,
2, Avenue Augustin Fresnel, 91127 Palaiseau Cedex, France}
\affiliation{Universit\'{e} de Sherbrooke, Department of Mechanical Engineering, Sherbrooke, PQ J1K 2R1, Canada.}

\author{Brahim Guizal}
\affiliation{Laboratoire Charles Coulomb (L2C), UMR 5221 CNRS-Universit\'{e} de Montpellier, F- 34095 Montpellier, France}

\author{Mauro Antezza}
\affiliation{Laboratoire Charles Coulomb (L2C), UMR 5221 CNRS-Universit\'{e} de Montpellier, F- 34095 Montpellier, France}
\affiliation{Institut Universitaire de France, 1 rue Descartes, F-75231 Paris Cedex 05, France}

\author{Svend-Age Biehs}
\affiliation{Institut f\"{u}r Physik, Carl von Ossietzky Universit\"{a}t, D-26111 Oldenburg, Germany}

\date{\today}


\begin{abstract}
Heat flux exchanged between two hot bodies at subwavelength separation distances can exceed the limit predicted by the blackbody theory. However this super-Planckian transfer is restricted  to these  separation distances. Here we demonstrate the possible existence of a  super-Planckian transfer at arbitrary large separation distances if the interacting bodies are connected in near-field with weakly dissipating hyperbolic waveguides. This result opens the way to long distance transport of near-field thermal energy.
\end{abstract}

\maketitle

\section{Introduction}

Since the pionneering work of Polder and van Hove~\cite{Polder} it is well known that the radiative flux exchanged between two hot bodies at subwavelength separation distances can exceed the limit predicted by the blackbody theory~\cite{Planck}, thanks to the extra contribution of evanescent waves. In presence of resonant surface modes such as surface plasmons or surface polaritons, collective electron or partial charge oscillations coupled to light waves at the surface, the radiative-heat exchange can even drastically surpass this limit~\cite{Rytovbook,Polder,Pendry1999,JoulainEtAl2005,Volokitin} by several orders of magnitude. In the last decade several limits for this enhancement effect were derived~\cite{Volokitin2004,BasuEtAl2009,JoulainPBA2010,BiehsEtAl2010,Biehs2012,MillerEtAl2015}. These discoveries have opened the way to promising technologies for near-field energy conversion~\cite{MatteoEtAl2001,NarayanaswamyChen2003}, data storage~\cite{Srituravanich} as well as active thermal management~\cite{inv_rev} at nanoscale with thermal rectifiers~\cite{OteyEtAl2010,BasuFrancoeur2011,vanZwol1,Iizuka2012,LiEtAl2013,PBA2013,Ito2014}, transistors~\cite{PBA_PRL2014,ItoEtAl2016}, memories~\cite{Slava,DyakovMemory}, and heat flux splitters~\cite{BiehsPBA2016} based on exchanges of evanescent thermal photons.

On the contrary, at long separation distance (i.e. in the far-field regime) the transfer of energy between two bodies out of thermal equilibrium results exclusively from propagating waves. If it is possible to extract the non-radiative waves, which are naturally confined on the surface of materials, using various diffraction mechanisms, the flux exchanged between two media cannot go beyond the Planck limit when the gap is filled by vacuum as it can be shown in the framework of the Landauer formalism~\cite{JoulainPBA2010,BiehsEtAl2010,BiehsPBA2016}, for instance.

The situation dramatically changes if a third body is introduced between the two reservoirs. The reason of this modification is twofold. First, the presence of a body modifies the optical properties of the medium between the two external bodies: thus, the evanscent waves existing at the interface between each reservoir and vacuum can be coupled to the third body and become propagating inside it. Secondly, the presence of a third body modifies in a more fundamental fashion the heat exchange, since the non-additivity of radiative heat transfer results in purely three-body effects that can hopefully be exploited to amplify the energy flux. This idea has been recently discussed in~\cite{Messina}, where the near-field radiative heat transfer between two bodies has been amplified thanks to the coupling of the reservoirs to a third thin slab placed between them.

\begin{figure}[hbt]
\includegraphics[scale=0.34]{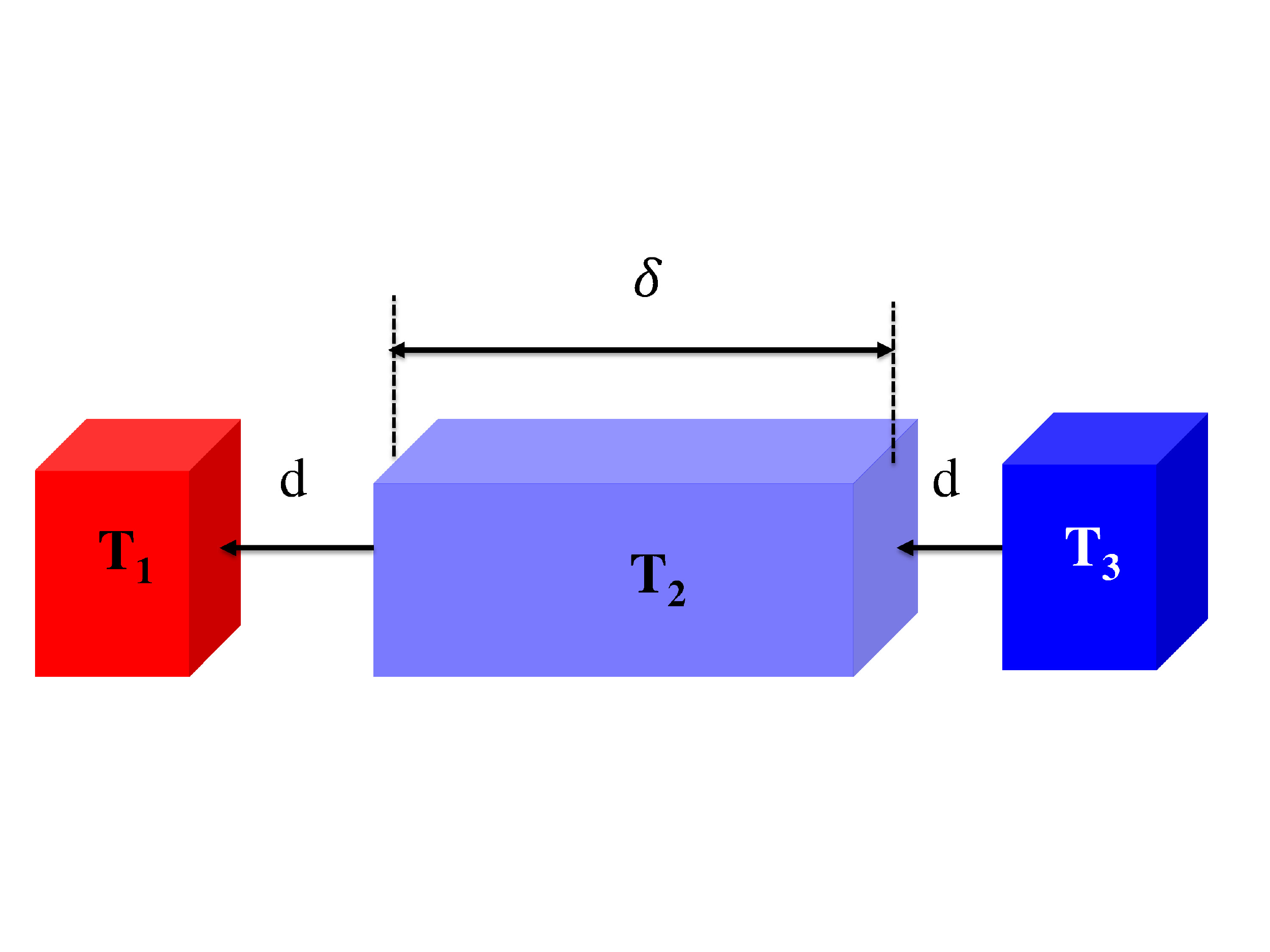}
\caption{Sketch of near-field heat pipe of length $\delta$ connecting a hot reservoir at temperature $T_1$ to a cold reservoir at temperature $T_3$. The separation distances between the pipe and the left and right reservoirs are equal to $d$.}
\label{waveguide}
\end{figure}

Differently from this last work, our attention is focused on the far field. We want to understand whether a third invervening body can be exploited to obtain a far-field heat transfer between two external planar slabs going beyond the blackbody limit. In particular, by choosing dielectric source and sink, we exploit the properties and the anisotropy of hyperbolic materials to produce a heat sink, transporting the near-field energy over distances larger than the thermal wavelength and going therefore beyond Planck's blackbody limit.

\section{Physical system}

Let us start our discussion with a review of the basics of the radiative heat flux between two semi-infinite reservoirs held at fixed temperatures $T_1$ and $T_3$, which are connected by thin slab of thickness $\delta$ and temperature $T_2$ as depicted in Fig.~\ref{waveguide}. Between both reservoirs and the intermediate slab is a vacuum gap of thickness $d$. According to the three-body theory of radiative heat transfer~\cite{Messina,Messina2}, which is based on Rytov's fluctuational electrodynamics~\cite{Rytovbook}, the heat flux $\Phi_3$ received in steady-state regime by the reservoir on the right side reads
\begin{equation}
\Phi_3 = \int_0^\infty\frac{d\omega}{2\pi} \hbar\omega \sum_{j = s,p}\int \frac{d^2 \kappa}{(2\pi)^2} \bigl[n_{12}\mathcal{T}^{12}_j+n_{23}\mathcal{T}^{23}_j \bigr],
\label{flux3b}
\end{equation}
where $n_{\alpha\beta}(\omega)=n_\alpha(\omega)-n_\beta(\omega)$, $n_\alpha(\omega)=(e^{\hbar\omega k_B T_\alpha}-1)^{-1}$ are the mean photon occupation numbers at equilibrium temperature $T_\alpha$ with $\alpha=\text{1},\text{2},\text{3}$. $\mathcal{T}^{\alpha\beta}_j(\omega,\mathbf\kappa)$ are the energy transmission coefficients for both polarizations $j = \rs,\rp$ which take into account the contributions of propagating ($\kappa<\omega/c$, $\kappa$ being the component of the wavevector parallel to the slabs) and evanescent waves in vacuum ($\kappa>\omega/c$). They are defined in terms of (optical) reflection and transmission coefficients of different media as
\begin{equation}
\begin{split} 
\mathcal{T}^{\text{12}}_j &= \begin{cases} {\displaystyle \frac{\left|\tau_{b,j}\right|^2 (1-\left|\rho_{1,j}\right|^2)(1-\left|\rho_{3,j}\right|^2) }{ \left|D_j^\text{123}D_j^\text{12}\right|^2}}, & \kappa < \frac{\omega}{c},\\
{\displaystyle \frac{4\left|\tau_{b,j}\right|^2 \Im\left(\rho_{1,j}\right)\Im\left(\rho_{3,j}\right)e^{- 4 \Im(k_z) d}}{ \left|D_j^\text{123}D_j^\text{12}\right|^2}}, & \kappa > \frac{\omega}{c}, \end{cases}\\
\mathcal{T}^{\text{23}}_j &= \begin{cases}{\displaystyle \frac{(1-\left|\rho_{12,j}\right|^2)(1-\left|\rho_{3,j}\right|^2) }{ \left|D_j^\text{123}\right|^2}}, & \kappa < \frac{\omega}{c},\\
{\displaystyle \frac{4 \Im\left(\rho_{12,j}\right)\Im\left(\rho_{3,j}\right)e^{- 2 \Im(k_z) d}}{ \left|D_j^\text{123}\right|^2}}, & \kappa > \frac{\omega}{c}.\end{cases}
\label{transmission_coefficients}
\end{split} 
\end{equation}
where $k_z=\sqrt{\omega^2/c^2-\kappa^2}$ is the normal component of the wavevector, while 
\begin{align}
D^\text{12}_j &= 1- \rho_{1,j}\rho_{b,j}e^{2i k_z d}, \\
D^\text{123}_j &= 1- \rho_{12,j}\rho_{3,j}e^{2i k_z d}
\end{align}
are the Fabry-P\'{e}rot-like denominators. Here $\rho_{1,j}$ and $\rho_{3,j}$ are the Fresnel reflection coefficients of the two reservoirs, while
\begin{align}
\tau_{b,j} &= \frac{(1 - \rho_{2,j}^2)\re^{i k_{z2} \delta}}{1 - \rho_{2,j}^2 \re^{2 i k_{z2} \delta}}, \\
\rho_{b,j} &= \rho_{2,j} \frac{1 - \re^{i 2 k_{z2} \delta}}{1 - \rho_{2,j}^2 \re^{2 i k_{z2} \delta}}
\end{align}
are the transmission and reflection coefficient of the intermediate body ($\rho_{2,j}$ being the corresponding Fresnel coefficient corresponding to a semi-infinite medium) and 
\begin{equation}
 \rho_{12,j} = \rho_{b,j} + \left(\tau_{b,j}\right)^2 \frac{\rho_{1,j}e^{2i k_z d}}{D_j^\text{12}}
\end{equation}
is the reflection coefficients of the left and intermediate bodies considered as a single entity. Here $k_{z2} = \sqrt{\omega^2/c^2 \epsilon_2-\kappa^2}$ is the normal component of the wave-vector inside the intermediate body.

It can be easily checked that if $\delta = 0$ then $\tau_{b,j} = 1$ and $\rho_{b,j} = 0$ so that $\rho_{12,j} = \rho_{1,j} \re^{2 i k_z d}$. When inserting these expressions the energy transmission coefficients reduce to the well-known expressions of Polder and van Hove~\cite{Polder} for the radiative heat flux between two semi-infinite planar reservoirs separated by a vacuum gap of distance $2d$. In this case (without intermediate slab), it is well-known that the radiative heat flux can be larger than that predicted by Stefan-Boltzmann law when the distance $2d$ becomes smaller than the thermal wavelength $\lambda_{\rm th}$ due to the extra contribution of evanescent waves~\cite{Polder,Pendry1999,Volokitin}. In many different experimental setups this super-Planckian radiation has been verified in the last ten years~\cite{KittelEtAl2005,HuEtAl2008,ShenEtAl2008,NatureEmmanuel,Ottens2011,Kralik2012,ShenEtAl2012,KimEtAl2015}. The near-field enhanced heat flux or super-Planckian radiation is particularly large if both reservoirs have surface phonon polariton resonances in the infrared~\cite{JoulainEtAl2005}. This is the case, for instance, for SiC and GaN, that we will use throughout the paper as examples of sources and sinks. As anticipated, our aim here is to study how the introduction of a hyperbolic intermediate slab can channel a super-Planckian radiative heat flux from reservoir 1 to reservoir 2 over distances which are larger than the thermal wavelength, of the order of $10\,\mu{\rm m}$ for $T = 300\,{\rm K}$.

To this end, we simplify the discussion by assuming that $T_1 = T + \Delta T$ and $T_2 = T_3 = T$. This corresponds to a situation where we start with the whole structure at equilibrium at temperature $T$ and then we heat up $T_1$ by the amount $\Delta T$. This results in a heat flux which is channeled towards reservoir 2. If the losses in the intermediate slab are small enough, then the temperature of the intermediate medium 2 will not change, which justifies the assumption that $T_2 = T_3 = T$. Under this assumption, the contribution proportional to $\mathcal{T}^{\text{23}}_j$ in Eq.~\eqref{flux3b} does not play any role, and we are led to define a heat-transfer cofficient as
\begin{equation}
\begin{split}
 H &:= \lim_{\Delta T \rightarrow 0} \frac{\Phi_3}{\Delta T} \\
 &:= \int_0^\infty\frac{d\omega}{2\pi} \frac{\rd \hbar \omega n(\omega)}{\rd T}\sum_{j = s,p}\int \frac{d^2 \kappa}{(2\pi)^2} \mathcal{T}^{12}_j.
\end{split}
\end{equation}
Here $n(\omega) = (\exp(\hbar \omega / \kb T) - 1)^{-1}$ is the mean occupation number of the thermal photons at temperature $T$. This expression, which is only valid for $\Delta T \ll T$, is much more compact than Eq.~(\ref{flux3b}) and it depends only on one temperature $T$. From this expression it becomes obvious that the heat transfer coefficient is quite sensitive with respect to the transmission coefficient $\tau_{b,j}$ of the intermediate slab. This is so, because $\tau_{b,j}$ determines the properties of the intermediate slab and in particular the eigenmodes inside the slab which can be used to guide or channel the radiative heat flux between both reservoirs. Since
\begin{equation}
 \tau_{b,j} \propto \re^{i k_{z2} \delta}
\end{equation} 
with $k_{z2} = \sqrt{\epsilon_2\omega^2/c^2 -\kappa^2}$ it is clear that waves with $\kappa > \sqrt{\epsilon_2} \omega/c$ are exponentially damped along the slab. For large $\delta$ such waves can therefore not be guided between both reservoirs. Therefore a material with a large permittivity $\epsilon_2$ would be ideal for the purpose of heat flux channeling, offering us a wide region of modes of the $(\kappa,\omega)$ plane being evanescent in vacuum, while propagating inside our waveguide.

\section{Isotropic waveguide}

To gain some insight into the mechanism we want to address, we start with the simple case of an isotropic waveguide. For this purpose we choose germanium, an excellent candidate for two reasons: in the infrared region it has a high dielectric permittivity ($\epsilon_{\rm Ge} = 16 \equiv \epsilon_2$) and negligible losses. It thus allows to have propagating modes with lateral wavevectors up to $\kappa^{\rm max} = \sqrt{\epsilon_{\rm Ge}} \omega/c = 4 \omega/c$ which means that waves with $\omega/c < \kappa < 4 \omega/c$ which are thermally excited in reservoir 1 can tunnel into the intermediate Ge slab (if $d$ is smaller than the thermal wavelength). Inside the Ge slab these waves are converted to propagating waves which can travel through the Ge slab until they reach the second vacuum gap where they can tunnel to the second reservoir.

\begin{figure}[hbt]
 \includegraphics[scale=0.23]{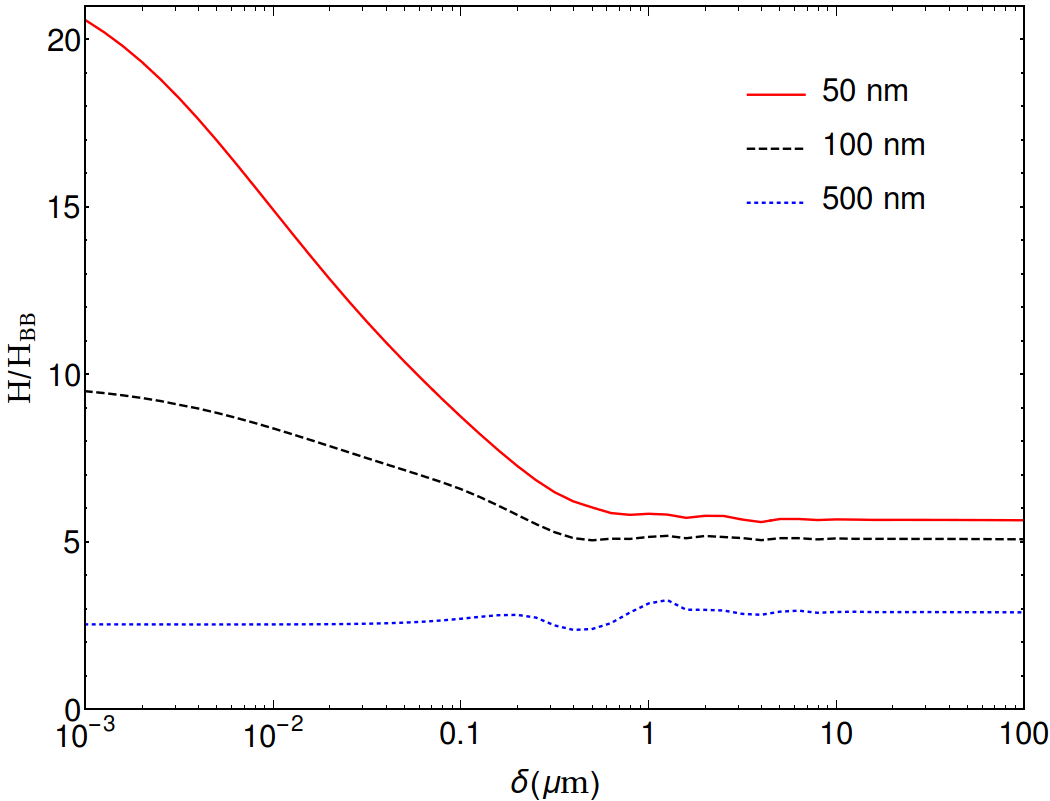}
 \caption{Heat transfer coefficient $H$ for SiC-Ge-SiC as function of the thickness of the intermediate germanium slab $\delta$ for a fixed gap distance of 
 $d = 50\,{\rm nm}$, $d = 100\,{\rm nm}$ and $d = 500\,{\rm nm}$. The heat-transfer coefficient is normalized to the blackbody value $H_{\rm BB} = 6.12 \, {\rm W} {\rm m}^{-2} {\rm K}^{-1}$ for $T = 300\,{\rm K}$.}
 \label{Fig:GeDelta}
\end{figure}

We show in Fig.~\ref{Fig:GeDelta} the heat transfer coefficient calculated for two SiC reservoirs, three different distances $d$ and as a function of the thickness $\delta$ of the intermediate slab. We describe the dielectric properties of SiC by means of a Drude-Lorentz model \cite{Palik98}
\begin{equation}\epsilon (\omega) =\epsilon_{\textrm{inf}}\frac{\omega ^2-\omega_\text{L}^2+i\gamma\omega}{\omega^2-\omega_\text{R}^2+i\gamma\omega},\end{equation}
with $\epsilon_{\infty}=6.7$, $\omega_{\rm L}=1.827\cdot 10^{14}\,$rad/s, $\omega_{\rm T}=1.495\cdot 10^{14}\,$rad/s and $\gamma=0.9\cdot 10^{12}\,$rad/s. We have considered the range of thicknesses $\delta \in [10^{-9}\,{\rm m},10^{-4}\,{\rm m}$]. We have plotted unphysical small values of $\delta = 10^{-9}\,$m just to illustrate the convergence. At such small $\delta$ the heat transfer coefficient $H$ converges to the value for two SiC reservoirs which are separated by a vacuum gap of thickness $2d$. In all the cases considered here ($d = 50,100,500\,$nm) this value is of course larger than the blackbody value $H_{\rm BB} = 6.12\,{\rm W}{\rm m}^{-2}{\rm K}^{-1}$, showing the super-Planckian effect. While this is not surprising for $\delta$ going to zero, since we fully are in a near-field regime, we have to focus on large values of $\delta$, and remark that for very large $\delta = 100\,\mu {\rm m}$ (i.e.\ $\delta \gg \lambda_{\rm th}$)
the heat transfer coefficient is still larger than the blackbody value although the overall distance is much larger than the thermal wavelength. This is exactly the waveguide effect we are looking for.

\begin{figure*}[!hbt]
 \subfigure[\,$|\tau_{2p}|$, $\delta = 10\,{\rm nm}$]{\includegraphics[scale=0.65]{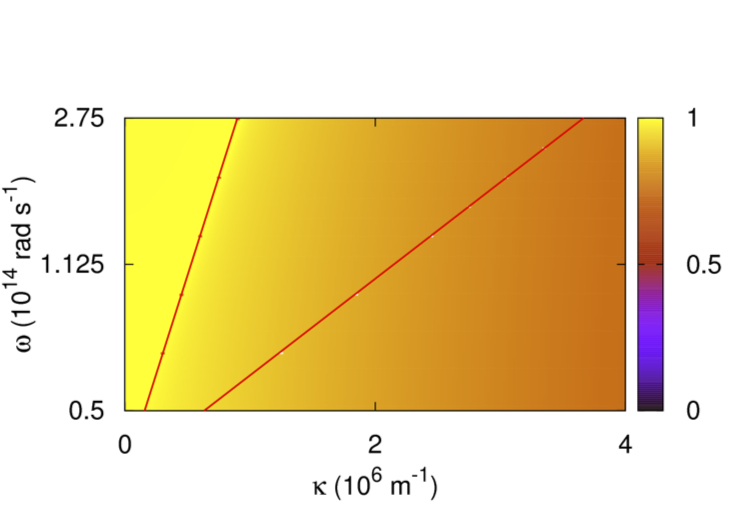}}
 \subfigure[\,$\mathcal{T}^{12}_\rp$, $\delta = 10\,{\rm nm}$]{\includegraphics[scale=0.65]{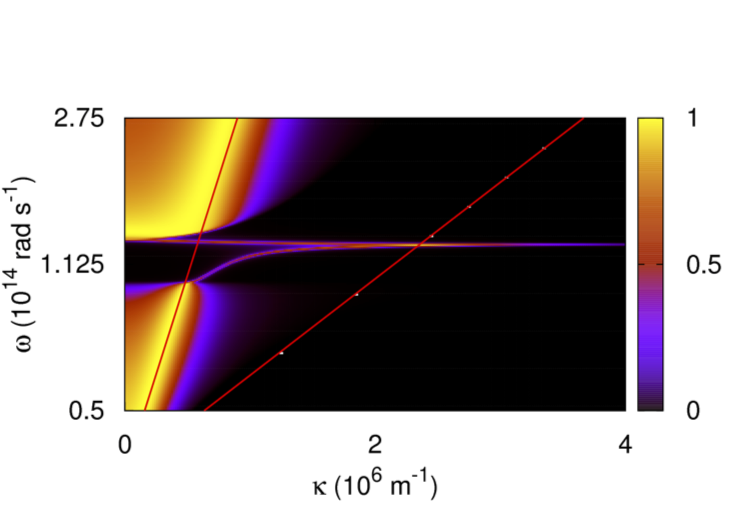}}
 \subfigure[\,$|\tau_{2p}|$, $\delta = 10\,\mu{\rm m}$]{\includegraphics[scale=0.65]{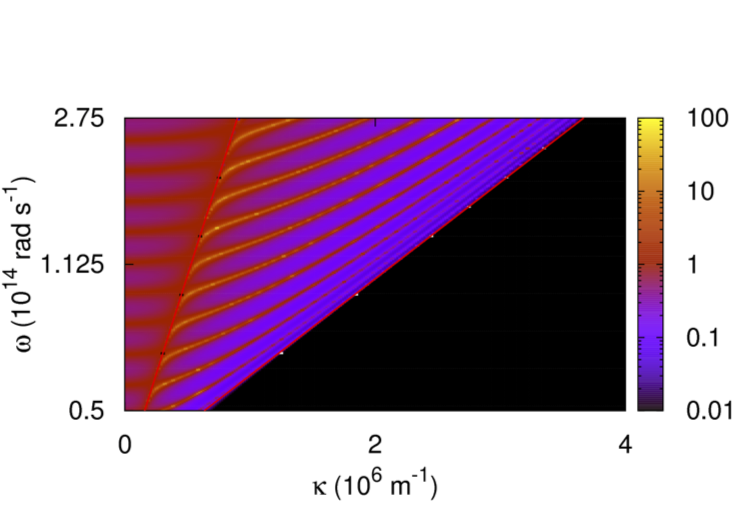}}
 \subfigure[\,$\mathcal{T}^{12}_\rp$, $\delta = 10\,\mu{\rm m}$]{\includegraphics[scale=0.65]{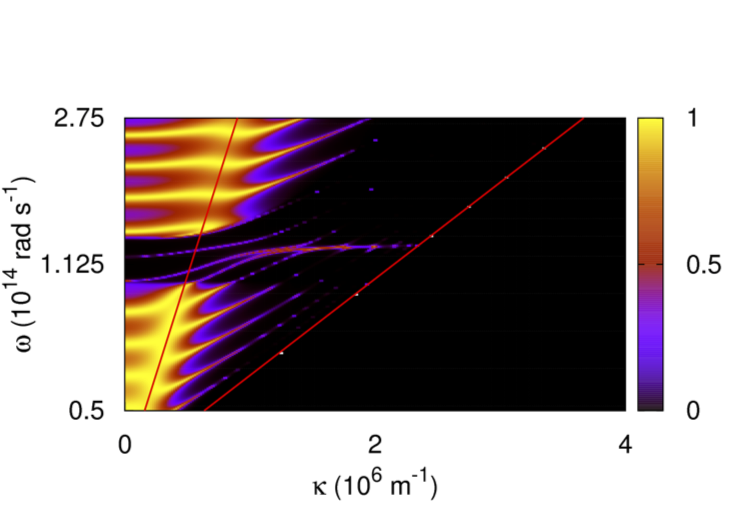}}
 \caption{Plot of the transmission coefficient of the intermediate slab $|\tau_{2p}|$ (left column) and the energy transmission $\mathcal{T}^{12}_\rp$ for p-polarized light for $d = 500\,{\rm nm}$ and different thicknesses $\delta$ of the intermediate slab in $(\kappa,\omega)$ space. The red lines are the light line in vacuum ($\omega = c \kappa$) and the light line in Germanium ($\omega = c \kappa/\sqrt{\epsilon_{\rm Ge}} $). As reservoir we use SiC.}
 \label{Fig:GeOmegaKappa}
\end{figure*}

The coupling of the surface polaritons inside the SiC reservoirs and the waveguide modes inside the Ge slab is shown in Fig.~\ref{Fig:GeOmegaKappa}. In the left column we have plotted the transmission coefficient of the intermediate slab $|\tau_{b,\rm p}|$ for the p polarization and in the right column the corresponding energy transmission of the
radiative heat flux $\mathcal{T}^{12}_\rp$, choosing $d = 500\,{\rm nm}$. For a very thin Ge slab with $\delta = 10\,{\rm nm}$ the transmission coefficient $|\tau_{b,\rm p}|^2$ is very close to 1 for all plotted $\omega$ and $\kappa$. Therefore all the waves in this $(\kappa,\omega)$ region are nearly perfectly transmitted. As a consequence, the energy transmission $\mathcal{T}^{12}_\rp$ plotted in Fig.~\ref{Fig:GeOmegaKappa}(b) is only slightly different from the energy transmission without the intermediate slab as it could be expected. In this plot the coupled surface phonon polaritons of both reservoirs can be very nicely seen. Now, when $\delta = 10\,\mu{\rm m}$ is very large, we find in Fig.~\ref{Fig:GeOmegaKappa}(c) that $|\tau_{b, \rp}|$ is very large between the light line in vacuum ($\omega = c \kappa$) and the light line in Ge ($\omega = c \kappa / 4$). It should be noted that $|\tau_{b,\rm p}|$ is much larger than 1 as a result of the poles (or better resonances) in $\tau_{b,\rm p}$ which are determined by the condition 
\begin{equation}
 1 = \rho_{2,j}^2 \re^{i 2 k_{z2} \delta}.
\end{equation}
These are the Fabry-P\'{e}rot modes inside the slab. The surface phonon polaritons as well as the total internal reflection modes of the reservoirs which are between the light lines in vacuum and in Ge can couple to these Fabry-P\'{e}rot modes, leading to a large transmission for these coupled modes as it can be seen in Fig.~\ref{Fig:GeOmegaKappa}(d). 

\section{Ideal anisotropic waveguide}\label{SecIdeal}

In the previous section we have seen that the super-Planckian heat radiation can be efficiently channeled or guided through an intermediate slab if this slab supports propagating waves with $\kappa > \omega/c$ in a broad frequency and wavevector range. As an alternative to the Ge slab we want to study now uni-axial slabs and in particular hyperbolic or indefinite materials~\cite{Smith2003}. These materials have already been considered for thermal radiation, because they allow for broad-band radiative heat fluxes~\cite{Nefedov2011,Biehs2012,GuoEtAl2012,Biehs2,ShiEtAl2015, LiuEtAl2014,LiuEtAl2015}, with a large penetration depth~\cite{LangEtAl2013,Tschikin2015} in contrast to phonon polaritonic materials~\cite{BasuZhang2009}. The advantage of hyperbolic materials is that in such materials propagating waves with large wavevectors can exist, a property that can be exploited for hyperbolic lensing~\cite{JacobEtAl2006,FengAndElson2006,CegliaEtAl2014}, for instance. Our goal is now to discuss this hyperbolic lensing for thermal radiation. Since we are looking for an ideal waveguide we will neglect losses (we actually introduce a very small imaginary part for each component of $\epsilon$, equal to $10^{-5}$) and dispersion during the discussion. For a real material one has to include losses as well as dispersion. Nonetheless such an idealization helps to find optimal parameters for the heat flux tunneling, which can serve as a basis for the search of a real hyperbolic waveguide structure.

Assuming that the optical axis of the uni-axial intermediate slab is along the $z$ direction, we can use the same heat flux expression as before since in this case there is no depolarization~\cite{YehBook,Bimonte,Opt_Exp,Messina2011}. We just need to replace the reflection coefficients $\rho_{b,j}$ by the corresponding uni-axial expressions~\cite{YehBook}
\begin{equation}
\begin{split}
 \rho_{b,\rs} &= \frac{k_z - k_{z,\rm o}}{k_z + k_{z,\rm o}}, \\ 
 \rho_{b,\rp} &= \frac{k_z \epsilon_\perp - k_{z,\rm e}}{k_z \epsilon_\perp + k_{z,\rm e}},
\end{split}
\end{equation}
with the wavenumbers $k_{z,\rm o}$ of the ordinary and $k_{z, \rm e}$ of the extra-ordinary waves fullfilling the dispersion relations~\cite{YehBook}
\begin{equation}
 \frac{k_{z,\rm o}^2}{\epsilon_\perp} + \frac{\kappa^2}{\epsilon_\perp} = \frac{\omega^2}{c^2},
 \quad\text{and} \quad \frac{k_{z,\rm e}^2}{\epsilon_\perp} + \frac{\kappa^2}{\epsilon_\parallel} = \frac{\omega^2}{c^2},
\end{equation}
where $\epsilon_\parallel$ ($\epsilon_\perp$) is the permittivity parallel (perpendicular) to the optical axis.

\begin{figure}[hbt]
\includegraphics[scale=0.7]{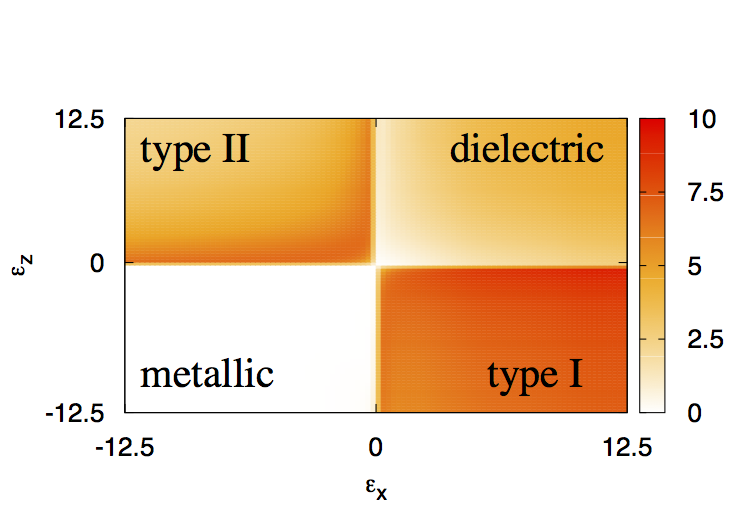}
 \caption{Heat-transfer coefficient $H$ for SiC-HM-SiC for different values of $\epsilon_x = \epsilon_\perp$ and $\epsilon_z = \epsilon_\parallel$, for $\delta = 10\,\mu{\rm m}$ and $d = 100\,{\rm nm}$. The HTC is normalized to the blackbody value $H_{\rm BB} = 6.12 \, {\rm W} {\rm m}^{-2} {\rm K}^{-1}$ for $T = 300\,{\rm K}$.}
 \label{Fig:HeatFluxMap}
\end{figure}

Furthermore, it is necessary to replace $k_{z2}$ in the expressions for s polarization by $k_{z,\rm o}$ and by $ k_{z,\rm e}$ for p polarization. Therefore, the transmission coefficient for p polarization of the intermediate slab satisfies
\begin{equation}
 \tau_{b,j} \propto \re^{i k_{z, \rm e} \delta}.
\end{equation}
This implies that only modes for which
\begin{equation}
 k_{z, \rm e} = \sqrt{\epsilon_\perp \frac{\omega^2}{c^2} - \kappa^2 \frac{\epsilon_\perp}{\epsilon_\parallel}}
\end{equation}
is a real number can efficiently guide the heat flux through the intermediate slab with large $\delta$. In order get some more insight we have to study the behavior of $k_{z, \rm e}$ in different regions of the plane $(\epsilon_\perp,\epsilon_\parallel)$: 
\begin{enumerate}
 \item $\epsilon_\perp > 0$ and $\epsilon_\parallel > 0$: In this case we have a dielectric uni-axial material. For such materials we have propagating waves in the slab for $\kappa < \sqrt{\epsilon_\parallel} \omega/c$. As a result, materials with large $\epsilon_\parallel$ are very useful for our purpose. 
 \item $\epsilon_\perp > 0$ and $\epsilon_\parallel < 0$: This is a so-called type-I hyperbolic material. In this material we can write the wavenumber of extra-ordinary waves as 
 \begin{equation} 
 k_{z, \rm e} = \sqrt{\epsilon_\perp \frac{\omega^2}{c^2} + \kappa^2 \frac{\epsilon_\perp}{|\epsilon_\parallel|}},
 \end{equation} 
 showing that for such materials $k_{z, \rm e}$ is real for all $\kappa$, meaning that such materials support propagating waves with arbitrarily large wavevectors. Of course in a real material there will be a cutoff at a given $\kappa$ which is determined by the microscopic properties of the structure. The type-I hyperbolic material is clearly an ideal candidate for our purpose. 
 \item $\epsilon_\perp < 0$ and $\epsilon_\parallel > 0$: This is a so-called type II hyperbolic material. In this case we can write the wavevector of extra-ordinary waves as 
 \begin{equation} 
 k_{z, \rm e} = \sqrt{- |\epsilon_\perp| \frac{\omega^2}{c^2} + \kappa^2 \frac{|\epsilon_\perp|}{\epsilon_\parallel}}.
 \end{equation} 
 Therefore $k_{z, \rm e}$ is real if $\kappa > \sqrt{\epsilon_\parallel} \omega/c$. Again we can have propagating waves inside the slab for arbitrary large $\kappa$, at least in principle. In contrast to the case of a dielectric uni-axial material, it is advantageous to have small $\epsilon_\parallel$ in this case.
 \item $\epsilon_\perp < 0$ and $\epsilon_\parallel < 0$: This anisotropic metallic case has to be treated separately. Actually in this case $k_{z,\rm e}$ is always imaginary. Therefore all the waves are damped inside the slab. But this is not the whole story. Actually it can be shown that for the special case where $\epsilon_\parallel \epsilon_\perp = 1$ the transmission coefficient of the intermediate slab is~\cite{HuChui2002}
 \begin{equation}
 \tau_{b,j} \propto \re^{\kappa \delta},
 \end{equation}
 in the quasistatic limit. This means that for $\epsilon_\parallel \epsilon_\perp = 1$ the evanescent waves are amplified. This is nothing else than the perfect lens effect found by Pendry~\cite{Pendry2000} for an isotropic metallic slab with $\epsilon_\parallel = \epsilon_\perp = -1$. The condition $\epsilon_\parallel \epsilon_\perp = 1$ is a 
 generalization to the uni-axial case. For the radiative heat flux this effect has been demonstrated in~\cite{Messina}, thus we will not follow this route further. Another reason is that this effect is quite sensitive to losses and metals are typically very lossy so that we cannot use that amplification effect for guiding heat radiation.
\end{enumerate}

In Fig.~\ref{Fig:HeatFluxMap}(a) we show $H/H_{\rm BB}$ for different combinations of $\epsilon_\parallel$ and $\epsilon_\perp$. First, we observe the existence of several regions where the ratio is larger than 1, i.e. where super-Planckian heat transfer can be indeed guided to far-field distances. It seems, in particular, that the type-I hyperbolic materials are quite advantageous for heat channeling because one can have large values of $H/H_{\rm BB}$ in a large parameter range of $\epsilon_\parallel$ and $\epsilon_\perp$. 

\begin{figure}[hbt]
\includegraphics[width=0.45\textwidth]{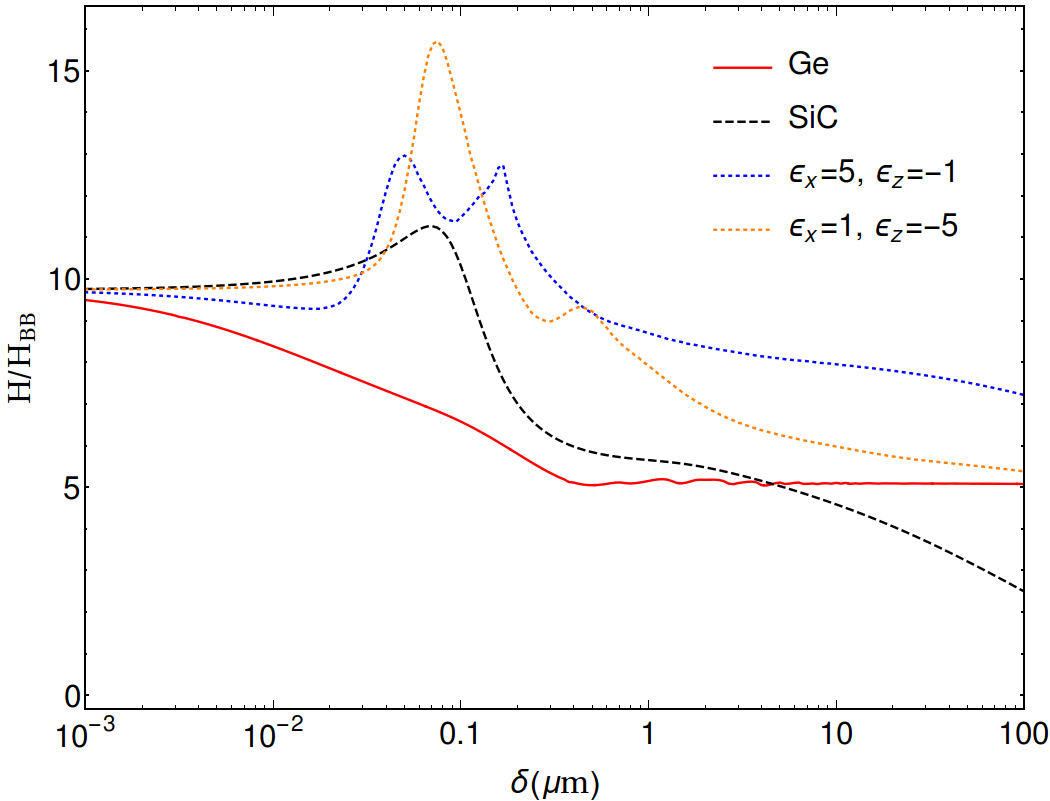}
 \caption{Heat-transfer-coefficient ratio $H/H_\text{BB}$ for SiC-HM-SiC as a function of the thickness $\delta$ of the intermediate slab for a fixed gap distance of $d = 100\,{\rm nm}$ and three different choices of $\epsilon_x = \epsilon_\perp$ and $\epsilon_z = \epsilon_\parallel$. The case of a SiC slab is also shown.}
\label{Fig:SlabthicknessTypeIandII}
\end{figure}

\begin{figure*}[hbt]
 \subfigure[\,$|\tau_{2p}|$, $\delta = 74\,{\rm nm}$]{\includegraphics[scale=0.65]{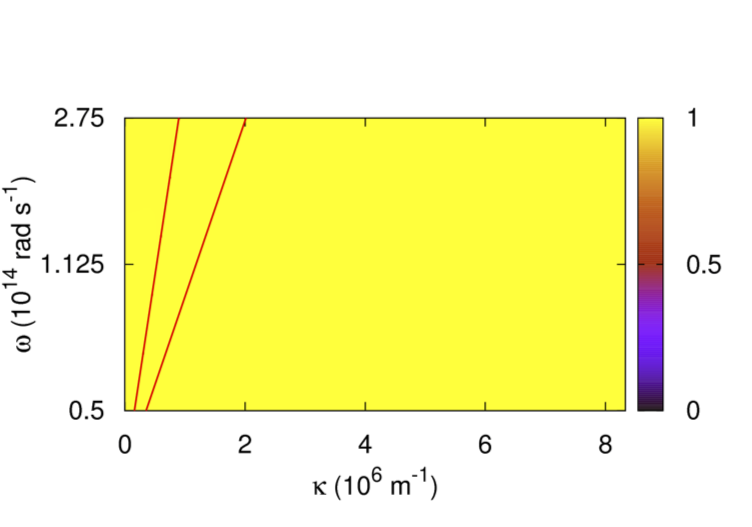}}
 \subfigure[\,$\mathcal{T}^{12}_\rp$, $\delta = 74\,{\rm nm}$]{\includegraphics[scale=0.65]{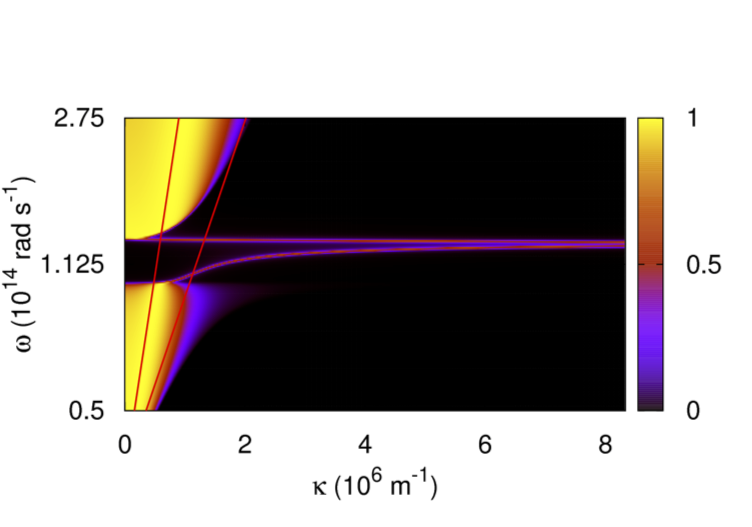}}
 \subfigure[\,$|\tau_{2p}|$, $\delta = 10\,\mu{\rm m}$]{\includegraphics[scale=0.65]{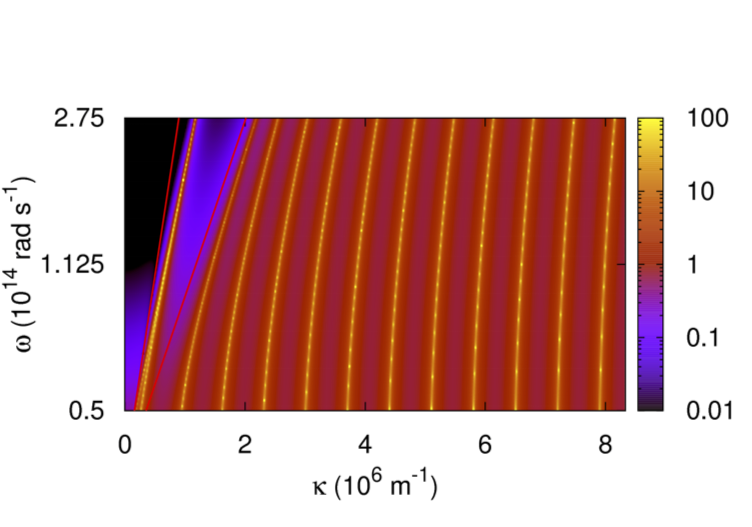}}
 \subfigure[\,$\mathcal{T}^{12}_\rp$, $\delta = 10\,\mu{\rm m}$]{\includegraphics[scale=0.65]{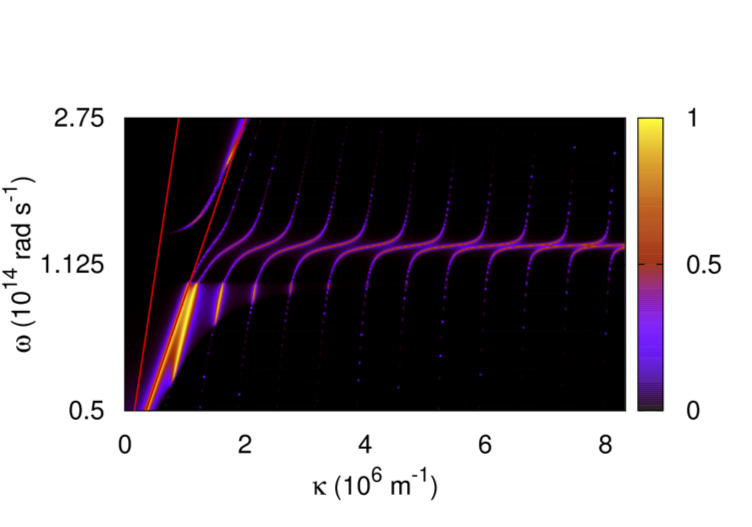}}
 \caption{Plot of the transmission coefficient of the intermediate slab $|\tau_{2p}|$ (left column) and the energy transmission $\mathcal{T}^{12}_\rp$ for p-polarized light for $d = 100\,{\rm nm}$ and different thicknesses $\delta$ of the intermediate hyperbolic slab of type II with $\epsilon_x = \epsilon_\perp = -1$ and $\epsilon_z = \epsilon_\parallel = 5$ in $(\omega$,$\kappa)$ space. The inserted lines are the light line in vacuum ($\omega = c \kappa$) and the light line $\omega = \sqrt{\epsilon_\parallel} c \kappa $). As reservoir we use SiC.}
 \label{Fig:T1}
\end{figure*}

The values of $H/H_{\rm BB}$ for three combinations of $\epsilon_\perp$ and $\epsilon_\parallel$ as a function of the slab thickness $\delta$ setting $d = 100\,{\rm nm}$ are shown in Fig.~\ref{Fig:SlabthicknessTypeIandII}. In particular, we consider the cases $(\epsilon_\perp,\epsilon_\parallel)=(-1,5)$ and $(\epsilon_\perp,\epsilon_\parallel)=(5,-1)$, corresponding respectively to a type-II and type-I hyperbolic material, the case $(\epsilon_\perp,\epsilon_\parallel)=(16,16)$ describing Ge, and compare these results to the case of using a SiC slab. In this plot it can be clearly seen that for large $\delta$  the type-I and type-II hyperbolic materials are better than Ge. Moreover, the type-I hyperbolic material is better than type-II for the chosen parameters. Interestingly for a thickness of $\delta\simeq100\,{\rm nm}$ we find values for $H/H_{\rm BB}$ which are larger than for $\delta \rightarrow 0$ indicating an enhancement or amplification effect for the hyperbolic materials (hyperlens effect) which does not exist for a Ge slab. This amplification is similar to that found in~\cite{Messina} (perfect-lens effect) where a thin metallic layer where used. Nevertheless, in this case the total distance $2d+\delta$ between the two reservoirs is still in the near-field regime.

\begin{figure*}[hbt]
 \subfigure[\,$|\tau_{2p}|$, $\delta = 10\,\mu{\rm m}$]{\includegraphics[scale=0.65]{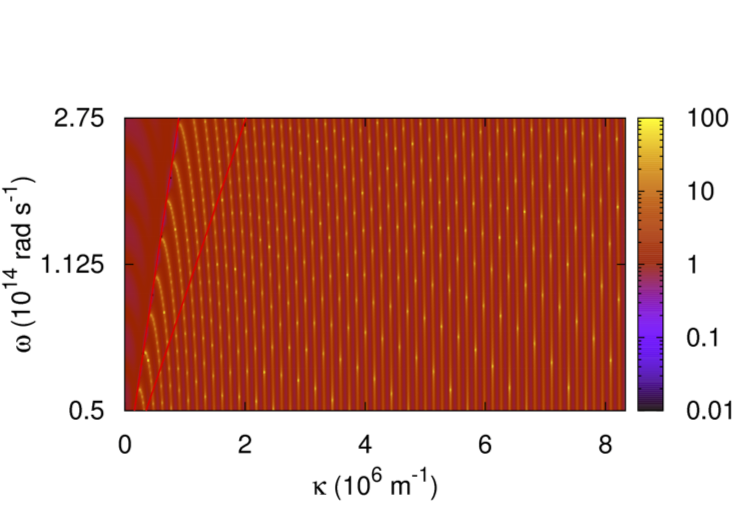}}
 \subfigure[\,$\mathcal{T}^{12}_\rp$, $\delta = 10\,\mu{\rm m}$]{\includegraphics[scale=0.65]{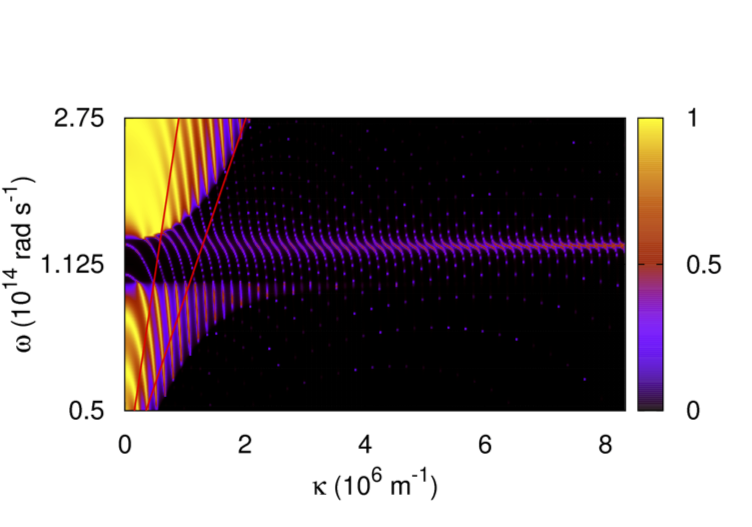}}
 \caption{Plot of the transmission coefficient of the intermediate slab $|\tau_{2p}|$ (left column) and the energy transmission $\mathcal{T}^{12}_\rp$ for p-polarized light for $d = 100\,{\rm nm}$ and different thicknesses $\delta$ of the intermediate hyperbolic slab of type I with $\epsilon_x = \epsilon_\perp = 5$ and $\epsilon_z = \epsilon_\parallel = -1$ in $(\omega$,$\kappa)$ space. The inserted lines are the light line in vacuum ($\omega = c \kappa$) and the light line $\omega = \sqrt{\epsilon_\parallel} c \kappa $). As reservoir we use again SiC.}
 \label{Fig:T2}
\end{figure*}

In Figs.~\ref{Fig:T1} and \ref{Fig:T2} we show the plots of $|\tau_{b,\rp}|$ and of $\mathcal{T}^{12}_\rp$ for the two hyperbolic materials. The coupling between the surface modes and the Fabry-P\'{e}rot modes in the hyperbolic slab can be nicely seen in both figures illustrating the channeling of the surface mode resonances through the hyperbolic waveguide. Note that the slope of the Fabry-P\'{e}rot modes in the type I hyperbolic materials is negative which is due to the fact that type-I hyperbolic materials show negative refraction~\cite{SmithEtAl2004,HoffmanEtAl2007}.

\begin{figure}[hbt]
\includegraphics[width=0.45\textwidth]{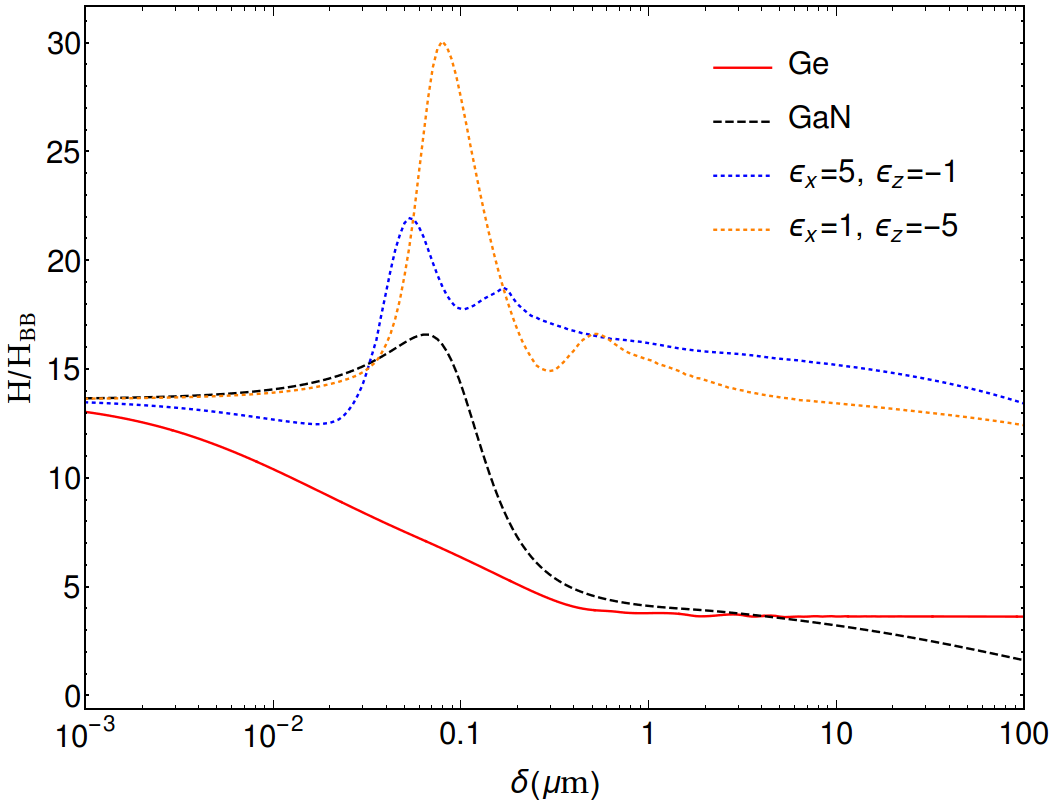}
 \caption{Heat-transfer-coefficient ratio $H/H_\text{BB}$ for GaN-HM-GaN as a function of the thickness $\delta$ of the intermediate slab for a fixed gap distance of $d = 100\,{\rm nm}$ and three different choices of $\epsilon_x = \epsilon_\perp$ and $\epsilon_z = \epsilon_\parallel$. The case of GaN is also shown.}
\label{Fig:GaN}
\end{figure}

In order to check the robustness of our results with respect to the choice of the reservoirs we calculate the same heat-transfer-coefficient ratio for two gallium nitride (GaN) reservoirs. The dielectric properties of GaN can again be safely described in the frequency region of interest by means of a Drude-Lorentz model~\cite{Palik98}, by choosing the parameters $\epsilon_{\infty}=6.7$, $\omega_{\rm L}=1.827\cdot 10^{14}\,$rad/s, $\omega_{\rm T}=1.495\cdot 10^{14}\,$rad/s and $\gamma=0.9\cdot 10^{12}\,$rad/s. We plot in Fig.~\ref{Fig:GaN} the quantity $H/H_\text{BB}$ as a function of the thickness $\delta$ for four different slabs: the two hyperbolic materials considered before, as well as Ge and GaN itself. As evident from the figure, apart from an overall increased value of the amplification factor, the same kind of conclusions can be drawn. First of all, also in this case in the limit of large thickness $\delta$, both hyperbolic materials perform well in guiding modes into the far field and are even better than Ge. Morevoer, we observe again the small-thickness peak in the amplification, and the prefernece of type-I over type-II in the regime of large thickness.

\section{Real anisotropic waveguide}

The considerations in the last section were made assuming the ideal case of a dispersion and dissipationless material. As it is well-known, causality demands both dispersion and dissipation which are connected by the Kramers-Kronig relations. Nonetheless, one can hope to find materials which fullfill the type-I hyperbolic property in the infrared with small dispersion and dissipation in the frequency window which is important for thermal radiation. In order to explore the possibility of a real anisotropic material we consider the very simple case of a multilayer structure where thin slabs (orthogonal to the $z$ axis) of SiC and Ge are periodically alternated. This structure is described by a filling factor $f$, associated to the fraction of SiC present in one period. The case $f=1$ ($f=0$) gives back a standard SiC (Ge) slab. By choosing two SiC reservoirs, we hope in this way to exploit both the presence of a phonon-polariton resonance in one of the two materials (SiC) constituting the intermediate slab which can couple to the resonances of the reservoirs, and the fact that this artificial material can produce anisotropy and a hyperbolic behavior.

\begin{figure}[hbt]
\includegraphics[width=0.45\textwidth]{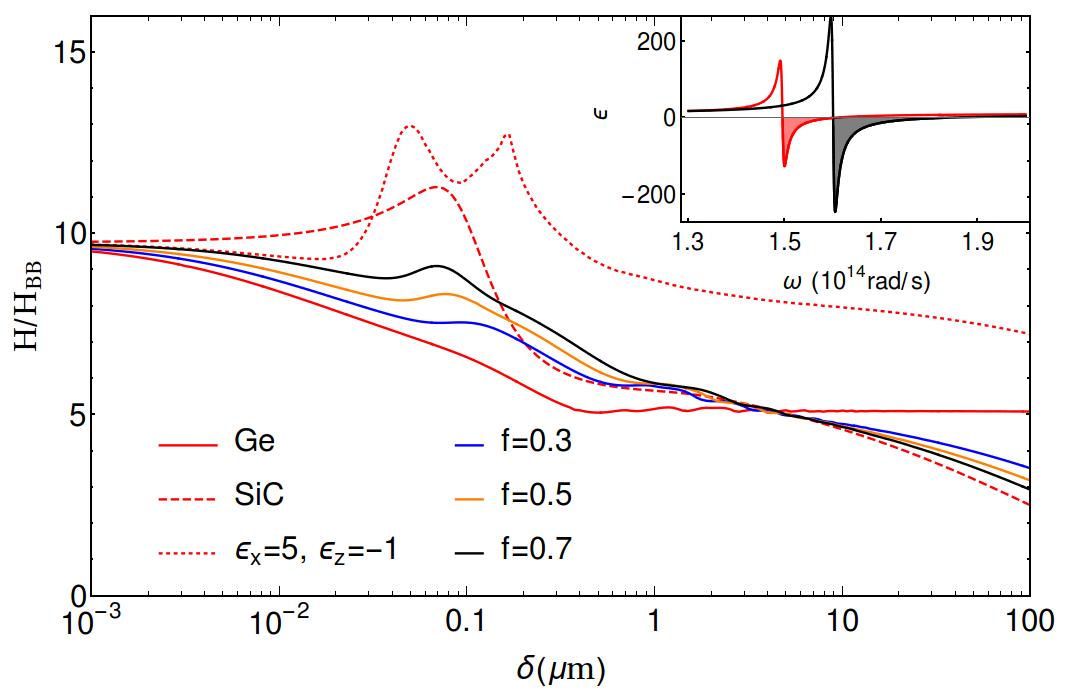}
 \caption{Heat-transfer-coefficient ratio $H/H_\text{BB}$ for SiC-HM-SiC. The hyperbolic material is a periodic arrangement of SiC and Ge thin films, having a SiC filling fraction $f$. The cases of SiC ($f=1$), Ge ($f=0$), and $(\epsilon_x,\epsilon_z)=(5,-1)$ are also shown for comparison. The inset shows $\epsilon_\perp$ (red line) and $\epsilon_\parallel$ (black line) for $f=0.5$, highlighting the hyperbolic regions (see text).}
\label{Fig:SiCGe}
\end{figure}

Before looking at the heat-transfer-coefficient ratio, let us check that this structure can indeed produce a hyperbolic behavior. To this aim we make us of the effective description in terms of an anisotropic dielectric permittivity. The perpendicular and parallel components of $\epsilon$ can be connected to the permittivities of SiC and Ge by means of the following expressions~\cite{Langhyp}, valid for each frequency:
\begin{equation}
 \begin{split}
  \epsilon_\perp&=f\epsilon_\text{SiC}+(1-f)\epsilon_\text{Ge},\\
  \epsilon_\parallel&=\frac{\epsilon_\text{SiC}\epsilon_\text{Ge}}{(1-f)\epsilon_\text{SiC}+f\epsilon_\text{Ge}}.
 \end{split}
\end{equation}
The result is plotted for a filling factor $f=0.5$ in the inset of Fig.~\ref{Fig:SiCGe}. It is clear that the two components $\epsilon_\perp$ (red line) and $\epsilon_\parallel$ (black line) have in some parts of the spectrum opposite sign. In particular we have a type-II hyperbolic behavior for lower frequencies (the red area in figure), while we have a type-I hyperbolic region for higher frequencies (the grey area).

Based on this observation we have calculated for three different filling factors the heat-transfer coeffiecient for this structure. The results are presented in the main part of Fig.~\ref{Fig:SiCGe}.

Several comments are in order. First of all, we notice that the behavior of this hyperbolic material is non-trivial, in the sense that the heat-transfer coefficient is not always intermediate between the one of SiC and the one of Ge. This can be clearly seen around $\delta\simeq400\,$nm. The heat-transfer coefficient becomes indeed intermediate between the ones of SiC and Ge for large values of $\delta$. Nevertheless, even if the final Super-Planckian amplification is lower than the one given by Ge alone, we stress the fact that by using a realistic material we still achieve our goal of channeling near-field effects to far field, since the ratio is still larger than 1.

The main message of the comparison between this analysis and the one performed in Sec.~\ref{SecIdeal} is that the presence of losses plays a key role in the existence and amplitude of this phenomenon. For this reason, we will in the next Section go back to the cases considered in Sec.~\ref{SecIdeal} and realize a quantitative study of the role of losses.

\section{The role of losses}

As anticipated, we perform in this Section a quantitative study of the role played by the losses of the intermediate slab. To this aim, we focus our attention on the case $(\epsilon_\perp,\epsilon_\parallel)=(5,-1)$ and introduce on both components of the anisotropic permittivity an imaginary part
\begin{equation}
\epsilon_\perp\to\epsilon_\perp+i\epsilon'',\quad\text{and}\quad\epsilon_\parallel\to\epsilon_\parallel+i\epsilon''.
\end{equation}

Also in this case, for the sake of simplicity, we assume this imaginary part $\epsilon''$ to be constant as well with respect to frequency.

\begin{figure}[hbt]
\includegraphics[width=0.45\textwidth]{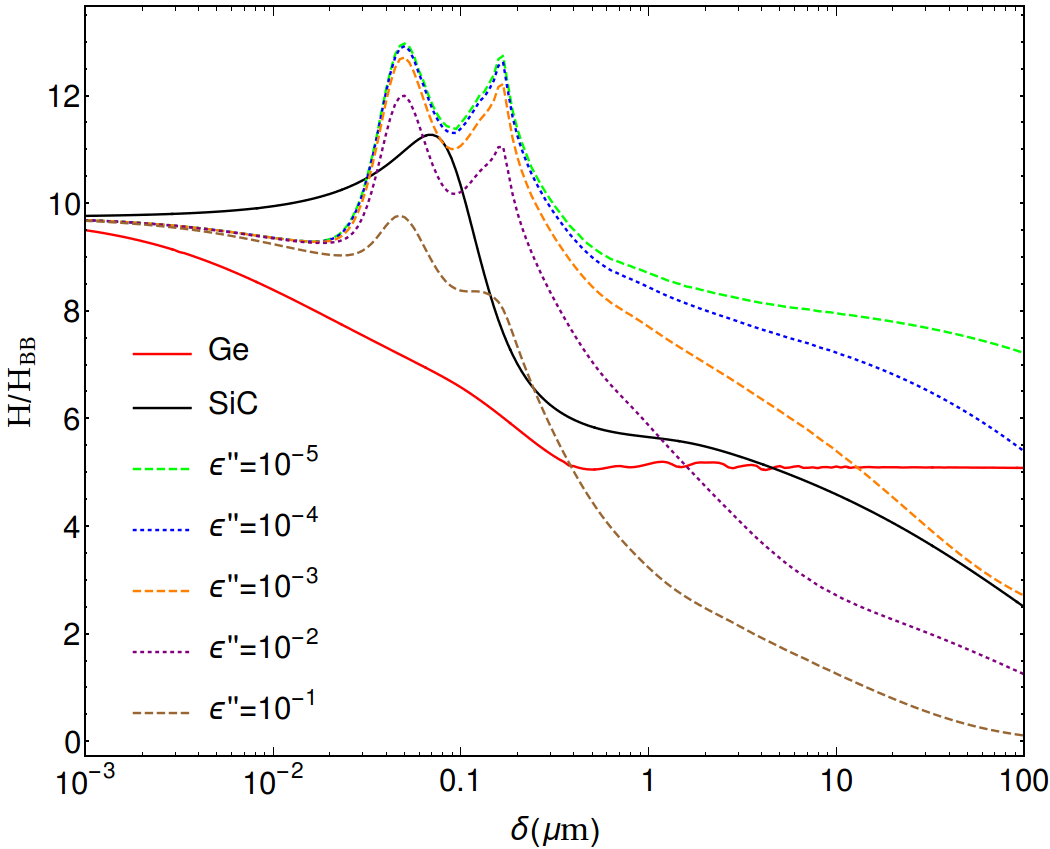}
 \caption{Heat-transfer-coefficient ratio $H/H_\text{BB}$ for the SiC-HM-SiC configuration. The solid lines correspond to the cases of SiC and Ge, while the dashed lines are associated with different choices of the imaginary part $\epsilon''$ of the permittivity of the intermediate slab.}
\label{Fig:LossesSiC}
\end{figure}

\begin{figure}[hbt]
\includegraphics[width=0.45\textwidth]{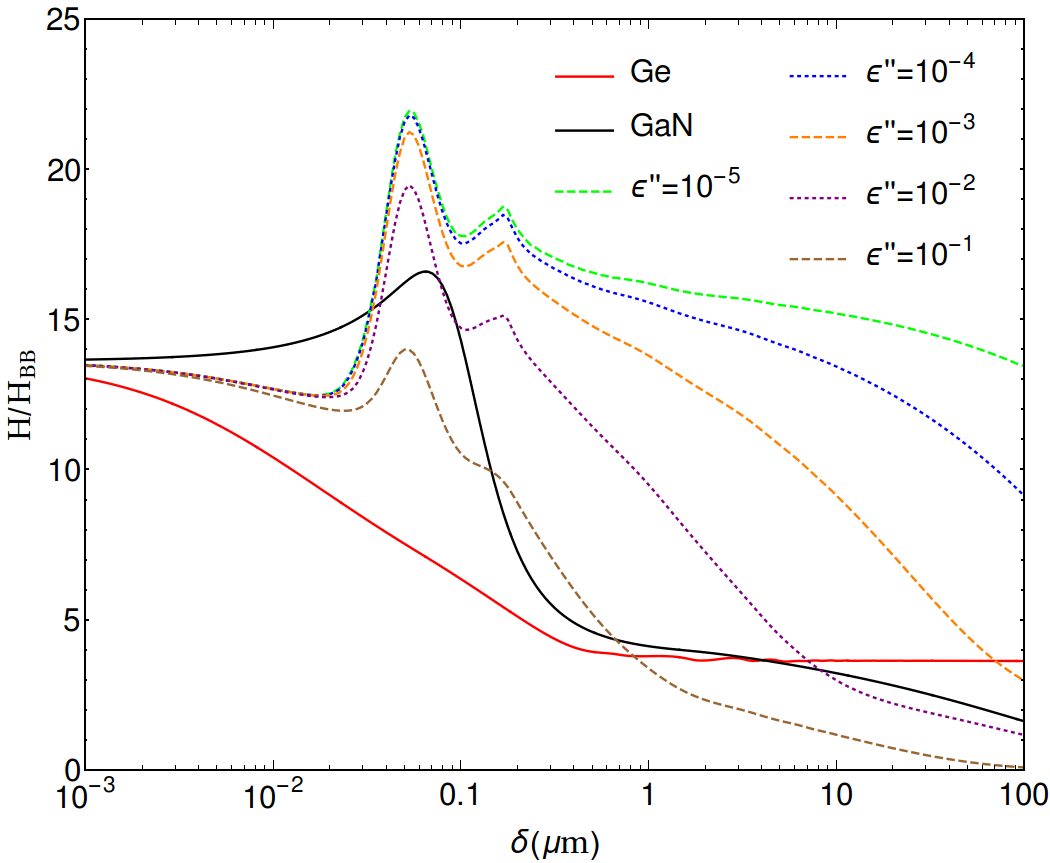}
 \caption{Heat-transfer-coefficient ratio $H/H_\text{BB}$ for the GaN-HM-GaN configuration. The solid lines correspond to the cases of GaN and Ge, while the dashed lines are associated with different choices of the imaginary part $\epsilon''$ of the permittivity of the intermediate slab.}
\label{Fig:LossesGaN}
\end{figure}

The results obtained are shown in Fig.~\ref{Fig:LossesSiC} in the case of SiC reservoirs. We show the case of imaginary part $\epsilon''=10^{-5}$, i.e. the one considered before, and go up to $10^{-1}$. We find that the ability of channeling Super-Planckian radiation on long distances is a very strong function of the losses in the intermediate material. This is not fully surprising, considering that we want to exploit propagating waves inside the middle slab, and the imaginary part $\epsilon''$ directly determines their typical decay length. We observe that for $\epsilon''=10^{-2}$ the effect is already for $\delta\simeq1\,\mu$m very important so that in this case the hyperbolic waveguide slab performs worse than both SiC and Ge for far-field distances. Nevertheless, even for $\epsilon''=10^{-2}$ we still have a ratio bigger than 1 for the largest $\delta$ considered here, whereas this is not the case for $\epsilon''=10^{-1}$. Finally, we show the same result in the case of two GaN reservoirs. Also in this case, we see basically the same trend and a strong transistion in the behavior of our hyperbolic waveguide happens around $\epsilon''=10^{-2}$.

\section{Conclusions}
In this work we have investigated heat exchanges by radiation between two hot bodies interconnected in near-field with anisotropic waveguides. We have predicted that a class of hyperbolic media could transport a super-Planckian heat flux over separation distances much larger than Wien's wavelength. By analyzing the transmission coefficients between these bodies we have shown that this behavior results from the presence of hyperbolic modes which remain propagating far beyond the light line. Hence, we have demonstrated that provided the exponential damping of these modes due to intrinsic losses of materials is weak the magnitude of heat flux exchanged between two hot bodies can be larger that that one predicted by Stefan-Boltzmann's law.

We believe that the hyperbolic waveguides could find broad applications in the field of thermal management by allowing the longdistance transport of the huge energy density which is usually confined close to the surface of materials. However, so far these waveguides have been considered as purely photonic systems. Further works are needed to evaluate the role play by the heat conduction on the heat transport.


\begin{thebibliography}{99}
\bibitem{Polder}D. Polder and M. Van Hove, Phys. Rev. B \textbf{4}, 3303 (1971).
\bibitem{Planck}M. Planck, The Theory of Heat Radiation (Dover, New York, 1991).
\bibitem{Rytovbook}S. M. Rytov, Y. A. Kravtsov, and V. I. Tatarskii, Principles of Statistical Radiophyics (Springer, New York), Vol. 3. (1989).
\bibitem{JoulainEtAl2005}K. Joulain, J.-P. Mulet, F. Marquier, R. Carminati, and J.-J. Greffet, Surface Science Report, \textbf{57}, 59-112 (2005).
\bibitem{Pendry1999}J. B. Pendry, Journal of Physics: Condensed Matter {\bf 11}, 6621 (1999).
\bibitem{Volokitin}A. I. Volokitin and B. N. J. Persson, Rev. Mod. Phys. \textbf{79}, 1291 (2007). 
\bibitem{Volokitin2004}A. I. Volokitin and B. N. J. Persson, Physical Review B {\bf 69} 045417(2004).
\bibitem{BasuEtAl2009}S. Basu and Z. M. Zhang, J. Appl. Phys. {\bf 105}, 093535 (2009).
\bibitem{MillerEtAl2015}O. D. Miller, S. G. Johnson, and A. W. Rodriguez, Phys. Rev. Lett. {\bf 115}, 204302 (2015).
\bibitem{JoulainPBA2010}P. Ben-Abdallah and K. Joulain, Phys. Rev. B \textbf{82}, 121419(R) (2010).
\bibitem{BiehsEtAl2010}S.-A. Biehs, E. Rousseau, and J.-J. Greffet, Phys. Rev. Lett. \textbf{105}, 234301 (2010).
\bibitem{Biehs2012}S.-A. Biehs, M. Tschikin, P. Ben-Abdallah, Phys. Rev. Lett. {\bf 109}, 104301 (2012).
\bibitem{MatteoEtAl2001}R. S. DiMatteo, P. Greiff, S. L. Finberg, K. A. Young-Waithe, H. K. Choy, M. M. Masaki, and C. G. Fonstad, Appl. Phys. Lett. \textbf{79}, 1894 (2001). 
\bibitem{NarayanaswamyChen2003}A. Narayanaswamy and G. Chen, Appl.Phys. Lett. \textbf{82}, 3544 (2003). 
\bibitem{Srituravanich}W. Srituravanich, N. Fang, C. Sun, Q. Luo, X. Zhang, Nano Lett. {\bf 4}, 1085-1088 (2004).
\bibitem{inv_rev}P. Ben-Abdallah and S.-A. Biehs, AIP Advances \textbf{5}, 053502 (2015).
\bibitem{OteyEtAl2010}C. R. Otey, W. T. Lau, and S. Fan, Phys. Rev. Lett. {\bf 104}, 154301 (2010). 
\bibitem{BasuFrancoeur2011}S. Basu and M. Francoeur, Appl. Phys. Lett. \textbf{98}, 113106 (2011).
\bibitem{vanZwol1}P. van Zwol, K. Joulain, P. Ben-Abdallah, and J. Chevrier, Phys. Rev. B, \textbf{84}, 161413(R) (2011).
\bibitem{Iizuka2012}H. Iizuka and A. Fan, J. Appl. Phys. {\bf 112}, 024304 (2012).
\bibitem{LiEtAl2013}J. Huang, Q. Li, Z. Zheng, and Y. Xuan, Int. J. Heat \& Mass Transf. {\bf 67}, 575 (2013).
\bibitem{PBA2013}P. Ben-Abdallah and S.-A. Biehs, Appl. Phys. Lett. {\bf 103}, 191907 (2013).
\bibitem{Ito2014}K. Ito, K. Nishikawa, H. Iizuka, and H. Toshiyoshi, Appl. Phys. Lett. {\bf 105}, 253503 (2014).
\bibitem{PBA_PRL2014}P. Ben-Abdallah and S.-A. Biehs, Phys.Rev. Lett. \textbf{112}, 044301 (2014).
\bibitem{ItoEtAl2016}K. Ito, K. Nishikawa, and H. Iizuka, Appl. Phys. Lett. {\bf 108}, 053507 (2016).
\bibitem{Slava}V. Kubytskyi, S.-A. Biehs and P. Ben-Abdallah, Phys. Rev. Lett. \textbf{113}, 074301 (2014).
\bibitem{DyakovMemory}S. A. Dyakov, J. Dai, M. Yan, and M. Qiu, arXiv:1408.5831 (2014).
\bibitem{BiehsPBA2016} S.-A. Biehs and P. Ben-Abdallah, Phys. Rev. B \textbf{93}, 165405 (2016).
\bibitem{Messina}R. Messina, M. Antezza, and P. Ben-Abdallah, Phys. Rev. Lett. {\bf 109}, 244302 (2012).
\bibitem{Messina2}R. Messina and M. Antezza, Phys. Rev. A \textbf{89}, 052104 (2014).
\bibitem{KittelEtAl2005}A. Kittel, W. M\"{u}ller-Hirsch, J. Parisi, S.-A. Biehs, D. Reddig, and M. Holthaus, Phys. Rev. Lett. {\bf 95}, 224301 (2005).
\bibitem{HuEtAl2008}L. Hu, A. Narayanaswamy, X. Chen, and G. Chen, \apl {\bf 92}, 133106 (2008).
\bibitem{ShenEtAl2008}S. Shen, A. Narayanaswamy, and G. Chen, Nano Lett. {\bf 9}, 2909 (2009).
\bibitem{NatureEmmanuel}E. Rousseau, A. Siria, G. Jourdan, S. Volz, F. Comin, J. Chevrier and J.-J. Greffet, Nature Photonics {\bf 3}, 514 (2009).
\bibitem{Ottens2011}R. S. Ottens \emph{et al.}, Phys. Rev. Lett. {\bf 107}, 014301 (2011).
\bibitem{Kralik2012}T. Kralik, P. Hanzelka, M. Zobac, V. Musilova, T. Fort, and M. Horak, Phys. Rev. Lett. {\bf 109}, 224302 (2012).
\bibitem{ShenEtAl2012}S. Shen, A. Mavrokefalos, P. Sambegoro, and G. Chen, App. Phys. Lett. {\bf 100}, 233114 (2012).
\bibitem{KimEtAl2015}K. Kim \emph{et al.}, Nature \textbf{528}, 387 (2015).
\bibitem{Palik98}\emph{Handbook of Optical Constants of Solids}, edited by E. Palik (Academic Press, New York, 1998).
\bibitem{Smith2003}D. R. Smith and D. Schurig, Phys. Rev. Lett. {\bf 90}, 077405 (2003).
\bibitem{Nefedov2011}I. S. Nefedov and C. R. Simovski, Phys. Rev. B {\bf 84}, 195459 (2011).
\bibitem{GuoEtAl2012}Y. Guo, C. L. Cortes, S. Molesky, and Z. Jacob, Appl. Phys. Lett. {\bf 101}, 131106 (2012).
\bibitem{Biehs2}S.-A. Biehs, M. Tschikin, R. Messina, and P. Ben-Abdallah, Appl. Phys. Lett. {\bf 102} 131106 (2013).
\bibitem{ShiEtAl2015}J. Shi, B. Liu, P. Li, L. Y. Ng, and S. Shen, Nano Letters {\bf 15}, 1217 (2015).
\bibitem{LiuEtAl2014}X. Liu, R. Z. Zhang, and Z. Zhang, ACS Photonics {\bf 1}, 785 (2014).
\bibitem{LiuEtAl2015}X. L. Liu and Z. M. Zhang, Appl. Phys. Lett. {\bf 107}, 143114 (2015).
\bibitem{LangEtAl2013}S. Lang, H. S. Lee, A. Y. Petrov, M. St\"{o}rmer, M. Ritter, and M. Eich, Appl. Phys. Lett. {\bf 103}, 21905 (2013).
\bibitem{Tschikin2015}M. Tschikin, S.-A. Biehs, P. Ben-Abdallah, S. Lang, A. Y. Petrov, and M. Eich, JQSRT {\bf 158}, 17 (2015).
\bibitem{BasuZhang2009}S. Basu and Z. M. Zhang, Appl. Phys. Lett. {\bf 95}, 133104 (2009).
\bibitem{JacobEtAl2006}Z. Jacob, L. V. Alekseyev, E. Narimanov, Opt. Exp. {\bf 14}, 8247 (2006).
\bibitem{FengAndElson2006}S. Feng and J. M. Elson, Opt. Exp. {\bf 14}, 216 (2006).
\bibitem{CegliaEtAl2014}D. de Ceglia, M. A. Vincenti, S. Campione, F. Capolino, J. W. Haus, and M. Scalora, Phys. Rev. B {\bf 89}, 075123 (2014).
\bibitem{Bimonte}G. Bimonte and E. Santamato, Phys. Rev. A {\bf 76}, 013810 (2007).
\bibitem{Opt_Exp}S.-A. Biehs, P. Ben-Abdallah, F. S. S. Rosa, K. Joulain, and J. J. Greffet, Optics Express, {\bf 19}, A1088 (2011).
\bibitem{Messina2011}R. Messina and M. Antezza, Phys. Rev. A {\bf 84}, 042102 (2011).
\bibitem{YehBook}P. Yeh, {\itshape Optical Waves in Layered Media}, (John Wiley \& Sons, New Jersey, 2005).
\bibitem{HuChui2002}L. Hu and S. T. Chui, Phys. Rev. B {\bf 66}, 085108 (2002).
\bibitem{Pendry2000}J. B. Pendry, Phys. Rev. Lett. {\bf 85}, 3966 (2000).
\bibitem{SmithEtAl2004}D. R. Smith, P. Kolinko, and D. Schurig, J. Opt. Soc. Am. B {\bf 21}, 1032 (2004).
\bibitem{HoffmanEtAl2007}A. J. Hoffman, L. Alekseyev, S. S. Howard, K. J. Franz, D. Wasserman, V. A. Podolskiy, E. E. Narimanov, S. L. Sivco, and C. Gmachl, Nature Mat. {\bf 6}, 946 (2007).
\bibitem{Langhyp}S. Lang, M. Tschikin, S.-A. Biehs, A. Y. Petrov, and M. Eich, Appl. Phys. Lett. \textbf{104}, 121903 (2014).
\end{thebibliography}
\end{document}